\documentclass[11pt]{article}
\usepackage{amssymb}
\newcommand{\comm}[1]{}

\def\short{\comm}

\usepackage{color}
\newcommand{\rd}[1]{\color{red}#1\color{black}} 
\newcommand{\gr}[1]{\color{green}#1\color{black}}

\newcommand{\gre}[1]{\color{magenta}#1\color{black}}

\usepackage{endnotes} 

\newtheorem{theorem}{Theorem}[section]
\newtheorem{proposition}[theorem]{Proposition}
\newtheorem{lemma}[theorem]{Lemma}
\newtheorem{corollary}[theorem]{Corollary}

\newtheorem{definition}[theorem]{Definition}
\newtheorem{remark}[theorem]{Remark}
\newtheorem{example}[theorem]{Example}

\def\e{\varepsilon}

\def\defi{\stackrel{{\scriptscriptstyle \Delta}}{=}}
\def\defi{:=}

\def\a{\alpha}
\def\d{\delta}
\def\o{\omega}
\def\O{\Omega}
\def\q{q}

\def\F{{\cal F}}
\def\w{\widehat}
\def \Ind{{\,\rm Ind\,}}
\def \Ind{{\mathbb{I}}}

\def\esssup{\mathop{\rm ess\, sup}}

\def\Re{{\rm Re\,}}

\def\R{{\bf R}}

\def\L{L}

\def\C{{\bf C}}

\def\ww{\widetilde}
\def\X{{\cal X}}

\def\oo{\bar}

\def\GG{{\cal G}}

\def\M{{\cal M}}

\def\L{{\cal L}}
\def\BB{{\cal C}}

\newcommand{\be}{\begin{equation}}
\newcommand{\ee}{\end{equation}}
\newcommand{\bd}{\begin{displaymath}}
\newcommand{\ed}{\end{displaymath}}
\newcommand{\ba}{\begin{array}{ll}}
\newcommand{\ea}{\end{array}}
\newcommand{\baa}{\begin{eqnarray}}
\newcommand{\eaa}{\end{eqnarray}}
\newcommand{\baaa}{\begin{eqnarray*}}
\newcommand{\eaaa}{\end{eqnarray*}}

\def\oo{\bar}

\def\a{\alpha}

\def\ew{\left(e^{i\o}\right)}

\def\ZZ{{\mathbb{Z}}}

\def\g{\gamma}
\def\G{\Gamma}
\def\ee{\e}

\def\AA{{\cal A}}
\def\BB{{{\cal B}(\rho)}}

\title{On  predictors and filters   for  non-decaying  unbounded continuous time  signals
}
\author{Nikolai Dokuchaev}
 
\begin{document}

 \def\short{\comm}
\def\brea{}
\def\breakk{}
\def\break{}
\def\BRR{}
\def\breakm{\nonumber\\  }\def\BR{}\def\BRR{}
\def\breacm{}
\def\dt{}
\def\ZZM{\ZZ_{\scriptscriptstyle{\le 0}}}

\def\BB{{\cal B}(\rho)}
\def\CC{{\cal C}}
\def\TP{{\frac{1}{2\pi}}}
\def\TPR{{\frac{\rho(t)}{2\pi}}}
\def\wH{{\w H}}
\def\Lrho{\rho^{-1}L_\infty(\R)}
\def\Linvrho {\rho^{} L_1(\R)}
\def\gre{}
\def\rd{\gr}
\def\gr{\comm}
\maketitle


 \begin{abstract}  
 The paper studies spectral representation and its applications for  non-decaying 
  continuous time signals
 that are not necessarily  bounded  at  $\pm\infty$. The paper introduces notions of transfer functions, spectrum degeneracy, spectrum gaps, 
 and  bandlimitness, for these unbounded signals. As an example of applications, 
explicit formulae are given  for transfer functions of  low-pass  and high-pass 
 filters  suitable for these signal.  As another example of applications, it is shown that 
 non-decaying unbounded signals with a single point spectrum degeneracy and sublinear rate of growth are predictable.
 The corresponding transfer functions for the  predictors are obtained explicitly.
 \end{abstract} 
\section{Introduction}
The paper studies continuous time  signals in a setting allowing non-decaying signals  such as signals featuring polynomial growth.
So far, unbounded functions of polynomial growth   have been studied by different special methods in the framework of the  approximation theory, see.e.g.
\cite{FeiMD,Han, NU}. However, these  signals are not sufficiently covered by the existing methods developed for  signal processing.   
 The purpose of this paper is to extend on the unbounded signals some  basic methods and results 
developed  for  signal processing.  

The most important  tools  used for signal processing  are based on the representation of signals
in the frequency domain. This includes, in particular,   criteria of predictability  and  data recoverability for signals. 
 In general, possibility of predicting or recovery  of  a signal from a sample is usually associated with restrictions on
the  class of underlying signals such as restrictions on the spectrum or sparsity.
For  continuous-time signals vanishing on $\pm \infty$, the predicting can be achieved for signals featuring some spectrum degeneracy. 
Technically, any   signal under some restrictions on its growth  can be modified  to a signal from $L_1(\R)$ without any loss of information, for example, by replacement $x(t)$ by $e^{-K|t|}x(t)$, for some larger enough $K>0$.  Similarly, an unbounded signal  also can be  can be transformed   to a signal from $L_1(\R)$ by multiplication 
on a damping coefficient, without any loss of information. 

 However,  at least for the case of  signals from $L_{1}(\R)$, these damping transformations represent  the convolutions in the frequency domain  
 with smoothing kernels. Unfortunately, 
these dumping transformations  would remove spectrum degeneracies  commonly exploited in data recovery and prediction for signal processing.  
For  unbounded two-sided signals  one could expect a similar impact of the damping transformations on the spectrum. 
So far, existing  standard methods based on spectral representations were not applicable to continuous non-decaying unbounded signals. 
Therefore, exploration of spectrum degeneracy  for non-decaying signals requires special consideration.

 For non-decaying bounded  signals with special type spectrum such as periodic and almost periodic signals, 
 there are Newton–Raphson-based algorithms that identifies the amplitudes, frequencies and phases
  of the sinusoidal components in the input signal to a high degree of accuracy; see, e.g. 
 \cite{Jog,So}. However, these methods are  not applicable for non-decaying signals with "continuous" spectrum, 
and are not applicable for unbounded signals in general. 

For the  purposes related to the problems of recoverability and prediction of digital signals, we need a straightforward 
 definition of spectral representations for  unbounded and non-decaying signals. 

Spectral representation for   general type non-decaying signals can be obtained via duality. For $p\in (1,\infty)$ and reflexive  weighted $L_p$-spaces of functions
 with polynomial growth these duality was used in (\cite{NU}) as  a tool for approximation by splines. 
For non-vanishing bounded  continuous time signal $x\in L_\infty(\R)$, i.e., from a non-reflexive space, a formal definition of the Fourier transform $X$ is given  in Chapter VI in \cite{Katz} via duality 
$\langle X,\oo f\rangle=\int_{-\infty}^\infty x(t)\oo F (t)dt $,  for   $f\in \AA$, where $\AA$ is the space of functions with Fourier transforms $F\in L_1(\R)$
(see also  \cite{FeiMD,NU}). 
In \cite{Kahane}, a similar approach is described for discrete time signals.  

The present paper considers a spectral representation  in non-reflexive  spaces for a wide  class of continuous time unbounded  signals 
represented as $\rho(t)^{-1}y(t)$, where $y\in L_\infty(\R)$ and $|\rho(t)|\to 0$ as $t\to\pm \infty$  (Section \ref{SecR}). 
This representation  is based on a  constructive  definition using    a limit of truncated sums, similarly to \cite{D24d}, 
where a similar definition  has been used therein for data recovery and predicting problems in discrete 
time setting.  

Based on representation, the paper extends the  notions of transfer functions, spectrum degeneracy, and band-limitiness,
 for these general type unbounded signals with a polynomial rate of growth
(Section  \ref{SecAp}).  
This also allowed to derive explicit formulae for low-pass and high-pass filters   for the case where $\rho(t)=(1+|t|)^{-\a}$ for $\a\ge 0$
 (Theorem \ref{ThFL} below).  
In addition, this allowed  to  establish 
predictability  and derive predictors for signals  
 with  a single point spectrum degeneracy and with sublinear rate of growth.
An explicit 
formula for the transfer function of the predictor is derived (Theorem \ref{ThP} below).    
Theorem \ref{ThP} represents a modification 
 of  Theorem 3.1
 from \cite{D21}, where standard  setting with signals decaying  on $\pm\infty$  has been considered.

\gr{It can be noted that the spectral representation introduced in Section \ref{SecR}  has been also  used in \cite{D24s} to obtain a sampling theorem and interpolation formula 
 on unbounded   band-limited  signals with sublinear rate of growth.  This interpolation formula 
 has   the coefficients that are decreasing with the rate $\sim 1/k$; this allowed to apply it for unbounded non-vanishing signals. }


  \subsection*{Some notations}

Let  $\R$ and $\C$, be the set of all real and complex numbers, respectively.

\def\DD{\mathbb{D}}

For an interval  $D\subseteq \R$, we denote by $ L_{\infty}(D) $ the set of all processes (signals) $x:D\to \C$, such that
$\|x\|_{ L_{\infty}(D) }\defi \esssup_{t\in D}|x(t)|<+\infty$.

For $r\in[1,\infty)$, we denote by $L_r(D)$ the set of all functions $x:D\to \C$, such that
$\|x\|_{L_r(D)}\defi \left(\int_{-\infty}^{\infty}|x(t)|^r dt\right)^{1/r}<+\infty$.

For $r\ge 1$, we denote by   $ W_r^1(D)$  the Banach  space of functions $f:  D \to\C$
that belong to $L_r(D) $ together with the distributional derivatives
up to the first order.

For $D=\R$ or a closed interval $D$, we denote by  $C(D)$  the standard linear  space of continuous functions $f:  D\to\C$
with the uniform norm $\|f\|_C\defi \sup_{t\in D} |f(t)|$.

\gr{For $r\in[1,\infty)$, we denote by $\ell_r$ the set of all processes (signals) $\xi:\ZZ\to \C$, such that
$\|\xi\|_{\ell_r}\defi \left(\sum_{k=-\infty}^{\infty}|\xi(k)|^r\right)^{1/r}<+\infty$. {We denote by $\ell_\infty$ the set of all processes (signals) $\xi:\ZZ\to \C$, such that
$\|\xi\|_{\ell_\infty}\defi \sup_{k\in\ZZ}|\xi(k)|<+\infty$.}}

\gr{Clearly, the embeddings 
$ W_r^1 (\R)\subset C( \R )  $ and 
$  C(\R)^*\subset W_r^1 (\R)^*$ are continuous for $r\ge 1$.}

We denote by $\Ind$  the indicator function. 
We denote by $\ast$ the convolution 
\baaa
(h\ast x)(t)\defi\int_{-\infty}^\infty h(t-s) x(s)ds, \quad t\in\R.
\eaaa

\section{Spectral representation of unbounded signals}
\label{SecR}

Let $\rho:\R\to (0,1]$ be a measurable function such that, for all $t,s\in\R$,
\baa\label{rho}
\rho(t)\rho(s)\le \rho(t+s).
\eaa

\gr{For $r\in[1,+\infty)$,  let $L_r(\R,\rho)$ be the space of functions $f:\R\to \C$ such that
$\rho f \in L_r(\R)$, 
with the  norm $\|f\|_{L_r(\R,\rho)}\defi \|\rho f\|_{L_r(\R)}$. }

Let $\Lrho$ be the space of functions $f:\R\to \C$ such that
$\rho f \in L_\infty(\R)$, 
with the  norm $\|f\|_{\Lrho}\defi \|\rho f\|_{L_\infty(\R)}$.

We consider below spectral representation and some applications for unbounded signals  from $\Lrho$.
It can be noted that  $\rho(t)=e^{-\a|t|}$ and $\rho(t)=(1+|t|)^{-\a}$ satisfy condition (\ref{rho}) for all $\a>0$.
  However, condition (\ref{rho})  does not hold, for example, for  $\rho(t)=e^{-|t|^2}$, so the corresponding class of signals $\Lrho$ is not covered.  

Let $\Linvrho$ be the space of functions $f:\R\to \C$ such that
$\rho^{-1} f \in L_1(\R)$, 
with the  norm $\|f\|_{\Linvrho}\defi \|\rho^{-1} f\|_{L_1(\R)}$. 

Let  $\BB$  be the space of functions $f:\R\to \C$ 
with the  finite norm $\|f\|_{\BB}\defi \|\w f(\o)\|_{\Linvrho}$, where 
$\w f(\o)=\int_{-\infty}^\infty e^{-i\o s}f(s)ds$ \comm{i.e., 
$\ww f(\o)=2\pi \w f(\o)$,  where $\w f(\o)=\frac{1}{2\pi}\int_{-\infty}^\infty e^{-i\o s}f(s)ds$}
 is the Fourier transform of $f$. Clearly, this class includes continuous functions only.
By the choice of its norm, $\BB$ is a separable Banach space that is isomorphic to $\Linvrho$.

We assume that each $X\in L_1( \R )$ represents an element of the dual space  $C( \R )^*$ such that
 $\langle X,f\rangle=\int_{-\infty}^\infty X(\o)f(\o)d\o$ for $f\in C( \R )$. We will use the same notation
  $\langle \cdot,\cdot\rangle$ for the extension of this bilinear form  
  on  $\BB^*\times \BB$.

 \begin{proposition}\label{prop1BB} 
\begin{enumerate}
\item 
If $f\in \BB$ and $g\in\BB$, then $h=fg\in\BB$ and $\|h\|_{\BB}\le \|f\|_{\BB}\|g\|_{\BB}$.  
\item Assume that $\rho(t)=(1+|t|)^{-\a}$, where $\a\ge 0$. Then, for any  integer
 $d>\a+1/2$ and any $r\in \Bigl(\max\left\{1,\frac{1}{d-\a}\right\},2\Bigr]$,  
   the embedding 
$ W_r^d (\R)\subset\BB$ is continuous  if $q>(k-\a)^{-1}$ for $q=1/(1-1/r)$.
\end{enumerate}
\end{proposition}
It can be noted that, since $d>\a+1/2$, we have that $1/(d-\a)\in (1,1/2)$.

\begin{proposition}\label{Prop1}  For any $x\in \Lrho$, there exists a 
weak* limit   $X\in \BB^*$ of the sequence of functions 
$X_m(\o)\defi \int_{-m}^m e^{-i\o t} x(t)dt$ defined on $ \R $ for $m>0$.
This $X$ is such that $\|X\|_{\BB^*}=\|x\|_{\Lrho}$.    
\end{proposition}

 In the lemma above,  $X_m\in \L_1(\R) \subset C( \R )^*\subset \BB^*$.

 We define a  mapping  $\F: \Lrho \to \BB^*$  such that  $X=\F x$ for $x\in  \Lrho$ is  the  limit  in $\BB^*$  introduced in Proposition \ref{Prop1}. By  Proposition \ref{Prop1}, this mapping is linear and continuous.

 Clearly, for $x\in L_1(\R)$, $\F x$ is the standard Fourier transform, and $\GG=\F^{-1}$ is 
 the inverse Fourier transform.

 \begin{lemma}\label{lemma2BB} 
 Let  $h\in \Linvrho$, and let $H$ be its Fourier transform.  Then $H e^{i\cdot t}=(h\ast e^{i\o \cdot})(t)$, 
 $H e^{i\cdot t}\in\BB$, and $\rho(t)\|H e^{i\cdot t} \|_{\BB}\le \|h\|_{\Linvrho}$  for any $t\in\R$. In addition, 
the function  $H e^{i\cdot t}$ is continuous in the topology of $\BB$ with respect to $t\in\R$.
\end{lemma}

\begin{proposition}\label{lemmahxy1} Let $h\in \Linvrho$.
 \begin{enumerate}
 \item For any  $x\in \Lrho$ and $X=\F x$, we have that 
 \baa\label{xX}
(h\ast x) (t) =\TP\langle X, H e^{i\cdot t} \rangle\quad \hbox{for}\quad t\in\R,
 \eaa
 and this convolution signal is continuous in $t$.
 \item For any $X\in\BB^*$, there exists an unique up to equivalency process 
 $x\in \Lrho$ such that (\ref{xX}) holds for any $h\in \BB$ for all $t$. 
 For this process, we have that $\|x\|_{\Lrho}\le \|X\|_{\BB^*}$,
 and $\F x=X$.
 \end{enumerate} 
\end{proposition}
\begin{remark}
 Clearly, for $x\in L_1(\R)$, $\F x$ is the standard Fourier transform, and $\GG=\F^{-1}$ is 
 the inverse Fourier transform. In this case, $X\in C(\R)$.   Since  $He^{i\cdot t}\in \Linvrho \subset L_1(\R)$,
 the right hand part of  (\ref{xX})  is defined by standard way as $\TP\int_{-\infty}^\infty  X(\o)H(\o)e^{i\o t}d\o$. 
 \end{remark}

We define an operator $\GG:\BB^*\to  \Lrho$ such that $x=\GG X$ in Proposition  \ref{lemmahxy1}(ii) above.

\begin{proposition}\label{lemmahxy} Let $h\in \Linvrho$.
 \begin{enumerate}
 \item For any  $x\in \Lrho$ and $X=\F x$, we have that $\rho(t)(h\ast x) (t)=y_h(t)$, where  $y_h(t)=\M_h(t) X$.
 \item For any $X\in\BB^*$ and  $y=\M_h X$, there exists an unique up to equivalency process 
 $x\in \Lrho$ such that $\rho(t)(h\ast x)(t)=y_h(t)$ for any $h\in \BB$ for all $t$. For this process, we have that $\|x\|_{\Lrho}\le \|X\|_{\BB^*}$,
 and $\F x=X$.
 \end{enumerate} 
\end{proposition}

\begin{theorem}\label{Th1} The mappings  $\F:\Lrho \to\BB^*$ and 
$\GG:\BB^*\to  \Lrho$ are continuous isometric bijections such that 
$\F=\GG^{-1}$ and  $\GG=\F^{-1}$. 
\end{theorem}
We write below $\F^{-1}$ instead of $\GG$.

The following proposition represents an analog of  Parseval's theorem for unbounded signals. 
\begin{corollary}\label{propP} For any $x\in \Lrho$ and $y\in \Linvrho$, we have that 
\baa
\int_{-\infty}^\infty x(t)\oo y(t)dt=\TP\langle X, \oo Y\rangle\quad \hbox{for}\quad X=\F x\in \BB^*, \quad Y=\F y\in \BB.
\eaa
\end{corollary}
\section{Spectrum degeneracy}
    The spectral representation introduced above allows to describe   signals from $\Lrho$ featuring spectrum degeneracy, 
    such as band limited processes and processes with spectrum vanishing at a single single point with certain rate. 
\subsection{Spectrum gaps and band-limitness}
\begin{definition}\label{defGap}
Let  $D\subset  \R $ be a Borel measurable set with non-empty interior,  and  let  
$x\in  \Lrho$ be such that  
 \baaa \int_{-\infty}^\infty x(t)f(t)dt=0 \eaaa
 for any $f\in \Linvrho$ such  that $F|_{ \R \setminus D}\equiv 0$ for $F=\F f$.
 In this case, we say that $D$ is a spectrum gap of $x$ and of $X=\F x$.
 \end{definition}
 
 \par
By Corollary \ref{propP}, a signal $x\in \Lrho$ has a spectrum gap $D\subset \R$ if and only if  
$\langle  \F x,F\rangle =0$
for any  $f\in \Linvrho$  such that  $F|_{ \R \setminus D}\equiv 0$, where $F=\F f$ is the Fourier transform of 
$f$.

In particular, if  $x\in \Lrho$ has a spectrum gap $D$,  and if  $f_1,f_2\in \BB$ are such that $f_1(\o)=f_2(\o)$ for all $\o\notin D$, 
then $\langle\F x,f_1\rangle = \langle\F x,f_2\rangle $.

As usual, we call a signal band-limited if it has a spectrum gap  $D$ such that the set $\R\setminus D$ is bounded.
\begin{example} Let $\O>0$ and an integer $m\ge 0$ be give. Then,  for any $\nu \in (-\O,\O)$,    the signal $t^k e^{i \nu t}$ belongs to $\Lrho$ with 
$\rho(t)=(1+|t|)^{-m}$ for  $k=1,2,...,m$, and the set $\R\setminus (-\O,\O)$ is its spectrum gap.  Respectively, any linear combination of such signals is a band-limited signal 
 with the frequency range  inside  $(-\O,\O)$.
\end{example}

\begin{remark} For the cases where 
$\rho(t)=e^{-\a|t|}$ or $\rho(t)=e^{-\a t^2}$  with $\a>0$, the concept of the spectrum gaps  in Definition \ref{defGap} 
is not useful since all functions  $f\in\BB$ are analytical for these $\rho$, and hence if $f_{\R\setminus D}\equiv 0$ then $f\equiv 0$. 
Hence  this definition implies that any set $D\subset \R$   with non-empty interior, and such that $D\neq \R$, is a spectrum gap for any signal
 from $\Lrho$ with these $\rho$.  \end{remark}

  The following theorem establishes that unbounded band-limited signals from $\Lrho$ featuring polynomial growth are  smooth.
  \begin{theorem}\label{ThCBL} Let $\rho(t)=(1+|t|)^{-\a}$ with $\a\ge 0$.
  In this case, for any $\O>0$, any  signal  $x\in\Lrho$ with  a spectrum gap $D=\R\setminus [-\O,\O]$  has continuous derivatives of any order. 
  \end{theorem}

\subsection{Spectrum degeneracy at a point}

\begin{definition}\label{defPointDeg}
Let $\G$ be a set.  Assume that a function $G:\R\times \G\to\C$ 
is such that $G(\cdot,\g)\in \C+\BB$ for each $\g\in \G$, and $\sup_{\g\in \G}\|G(\cdot,\g)\|_{\BB}= +\infty$.
Let $x\in \Lrho$ be such that $\sup_{\g\in \G}\|G(\cdot,\g)X\|_{{\BB^*}}<+\infty$.  
Then we say that the signal $x$ features spectrum degeneracy  that compensates   $G$.
\end{definition}
\begin{example}\label{ex1} Let $\rho(t)=(1+|t|)^{-\a}$ for $\a\in (0,1/2)$. Let $\G=(0,1)$, and  let 
$G(\o,\g)= (|\o|+\g)^{-1}$, $\nu\in\G$. Then  $G(\cdot,\g)\in W_r^1(\R)\subset \BB$ for any  $r>1$ and $\g\in\G$. 
Hence $G(\cdot,\g)\in \BB$.
 The corresponding spectrum degeneracy is a single point spectrum degeneracy. 
 \end{example}
\begin{example}\label{ex2}  Let $m\ge 1$ be a integer, let $\w\o\in\R$, and let $\rho(t)=(1+|t|)^{-\a}$ for $\a\ge 0$.
  Let $\G=(0,1)$,  and let
$G(\o,\g)= \exp\bigl(1/[(\o-\w\o)^{2m}+\g]\bigr)$. 
Then  $G(\cdot,\g)-1\in W_r^d(\R)\subset \BB$ for any  $r>1$ and any integer $d\ge 0$.  
 Hence $G(\cdot,\g)-1\in \BB$.
 The corresponding spectrum degeneracy is a single point spectrum degeneracy. 

 \end{example}
 
 \section{Transfer functions}
 \label{SecAp}

The transfer functions are commonly used in signal processing for signals $x(t)$ vanishing as  $t\to \pm\infty$, such as signals from
$L_p(\R)$ for $p=1,2$, and for signals that vanish either for all $t>0$ or for all $t<0$.
In particular, for any $ h \in L_1(\R)$, and its Fourier transform  $H=\F h$, we have that if $x\in L_1(\R)\cup L_2(\R)$ and $y=h\ast x$, then  $Y=\F y=HX$.  
However, the existing theory does not consider  transfer functions applied to the 
general unbounded signals $x(t)$ from  $\Lrho$ that do not vanish  as $t\to \pm\infty$.  
The suggested above spectral representation of  $x\in\Lrho$  via  $X=\F x\in \BB^*$ 
allows to implement the transfer functions from $\BB$ for  these signals. 
\subsection{Transfer functions for non-decaying unbounded signals}
 
\def\CBB{\C+\BB}

  We denote by $\CBB$ the set of all functions $G=a+H$, where $ a\in\C$ and $H\in\BB$.
  For  $X\in\BB^*$, $a\in\C$  and $H\in \BB$,  
  \begin{definition}\label{defTF}  We call
 $G\in \CBB$ a transfer function on $\Lrho $.  
\end{definition} 

 \begin{definition}\label{defTF}  We denote by $HX$ an element of $G\in\BB^*$ defined such that, for
  $\langle GX,f\rangle = \langle X,H f\rangle $ for any $f\in\BB$.
\end{definition}

By this definition, $\CBB$ is the set of all transfer function on $\Lrho$.

By Proposition \ref{Prop1}(i),   for any $f\in\BB$, we have that 
   $H f\in\BB$ as well. Hence if  $X\in\BB^*$,  $H\in \BB$,  and $Y=GX=aX+HX$, then $Y\in\BB^*$.

\begin{proposition}\label{propTF}  For $X\in\BB^*$,  $H\in \BB$,  
 $Y=HX\in \BB^*$,   we have that $y=h\ast x$.
\end{proposition}

\comm{  In addition, we have that
  $\langle Y,f\rangle = a \langle X,f\rangle+\langle X,H f\rangle $ for all $f\in\BB$.
  i.e.$ Y=a\d(\cdot)+ H X\in \BB^*$, where $\d(\cdot)\in C(\R)^*$ is a delta-function.}

This ensures that, in the definition above, $Y$ belongs to $\BB^*$ for $X\in \BB^*$.

 \begin{remark} In particular, the definition of the operator $\GG =\F^{-1}:\BB\to \Lrho$
 uses transfer functions  $H\in\BB$.  In the notations of Definition \ref{defTF}, we have that $y=h\ast x= \M_h X=F^{-1}Y=\F^{-1}  HX$.
 \end{remark} 

  \subsection{Causal and anti-causal transfer functions}
\begin{definition}\label{defCas}
We  say that a transfer function $H\in\CBB$ such as described in Definition \ref{defTF} is causal (or anti-causal) if, for any $\tau\in\R$ 
and any $x\in\Lrho$, if
$x(t)=0$ for a.e. $t<\tau$ ($x(t)=0$ for a.e. $t>\tau$), then $\w x(t)=0$ for all $t\le \tau$
 (or $x(t)=0$ for a.e. $t\ge\tau$, respectively).
 Here $X=\F x$, $\w  X=H X$, and $\w x=\F^{-1} \w X$. 
 \end{definition}
It can be noted that, in the definition above, $X,\w  X\in \BB^*$. 

Let $\C^-\defi \{z\in\C:\ \Re z< 0\}$, $\oo\C^-\defi \{z\in\C:\ \Re z\le 0\}$, 
$\C^+\defi \{z\in\C:\ \Re  z > 0\}$, and $\oo\C^+\defi \{z\in\C:\ \Re z\ge 0\}.$
\begin{theorem}\label{ThC}
\begin{enumerate}\item
Assume that a function $H:\oo\C^-\to\C$ 
is such that 
$H|_{\R}\in \CBB$. Assume that $H$ is continuous on  
 $\oo\C^-$ and  analytic on the interior of $\C^-$.
Then the trace of $H(\cdot/i)$ of $H$ on $\R$  
is a causal transfer function on $\Lrho$  in the sense of Definition  \ref{defCas}.
\item
Assume that a function $H:\oo\C^+\to\C$ 
is such that 
$H|_{\R}\in \CBB$. Assume that $H$ is continuous on  
$\oo\C^+$ and  analytic on the interior of $\C^+$.
Then this $H(\cdot/i)$ 
is an anti-causal transfer function on $\Lrho$  in the sense of Definition  \ref{defCas}.
\end{enumerate}
  \end{theorem}

 \section{Applications to signal processing}

In this section, we assume that $\rho(t)=(1+|t|)^{-\a}$, where $\a\ge 0$.

 \subsection{Low-pass and high-pass  filters}
Unfortunately, the ideal low-pass and high-pass filters with rectangle profile of the
 transfer function do not belong to $\CBB$. Hence they are not covered by 
Definition \ref{defTF} of the transfer functions  applicable to signals from $\Lrho$.   
However, 
 approximations of these ideal filters can be achieved  with  response functions from $\BB$. 

The following theorems provide examples of low-pass and high-pass filters applicable to non-decaying unbounded signals.

\begin{theorem}\label{ThFL}
Assume that $\rho$, $d$, and $r$,  are such as described in Proposition \ref{prop1BB} (ii). Let $0<p<q<\pi$, and 
let a  function $\mu:[p,q]\to\R$ be such that \baaa
\mu\in W_{r}^d([p,q]), \quad \mu(p)=1,\quad \mu(q)=0, 
\eaaa
and, if $d>1$, such that 
\baaa
\quad \frac{d^k\mu}{d\o^k}(\o)|_{\o=p+}=0,\quad \frac{d^k\mu}{d\o^k}(\o)|_{\o=q-}=0,\quad k=1,...,d-1.
\eaaa
In this case, the following holds.
\begin{itemize} \item[(i)]
 The function
\baa
H^{LP}_{p,q}\left(\o\right)\defi \Ind_{\{\o\in [-p,p]\}}+
\mu(\o)\Ind_{\{\o\in (p,q]\}} 
+\mu(-\o)\Ind_{\{\o\in [-q,-p)\}}
\label{LPF}\eaa
 is the transfer function for a low-pass filter from $\BB$, i.e. for any  $x\in \Lrho$, the signal $\w x=\F^{-1}(H^{LP}X)$ is a band-limited signal 
 with  a spectrum gap 
 $\R\setminus [-q,q]$
 \item[(ii)] The function \baaa
H^{HP}_{p,q}\left(\o\right)\defi 1-H^{LP}_{p,q}(\o),\quad 0<p<q<+\infty.
\eaaa
is the transfer function for a high-pass filter from $\CBB$, i.e. for any  $x\in \Lrho$, the signal $\ww x=\F^{-1}(H^{HP}X)$ is a signal 
 with  a spectrum gap 
 $ [-p,p]$.
 \item[(iii)]  If $\a<1/2$, then one can select $d=1$ and $\mu(\o)=\frac{q-\o}{q-p}$. 
  \item[(iv)]  If $\a<3/2$, then one can select $d=2$ and  \baaa
   \mu(\o)=1-\frac{1}{M}\left[ \frac{\o^3-p^3}{3}-\frac{p+q}{2}(\o^2-p^2)+qp(\o-p)\right], 
   \eaaa 
   where $M\defi\frac{\q^3-p^3}{3}-\frac{p+q}{2}(\q^2-p^2)+qp(q-p)$.
 \end{itemize}
 \end{theorem}
 \par
  For $q\to p+$, these functions  $H^{HP}_{p,q}$ and $H^{HP}_{p,q}$ approximate the ideal  low pass and high pass filters with  with the rectangular transfer functions
    $H^{LP}_{p,p}(\o)=\Ind_{\{\o\in [-p,p]\}}$ and $H^{HP}_{p,p}(\o)=1-\Ind_{\{\o\in [-p,p]\}}$.
    These step functions do not belong to $\CBB$.  Therefore, 
     these ideal filters  are not applicable to  signals from
     $\Lrho$.  In particular, the corresponding convolution integrals in time domain may diverge.
\par
It can be noted that   $\a\ge 1/2$ and $\rho$ is  such as described in  Proposition \ref{prop1BB} (ii)  
 we cannot claim that the piecewise linear functions  $H^{LP}_{p,q}$ and  $H^{HP}_{p,q}$ and  defined by $\mu$  as in Theorem \ref{ThFL}(iii) 
 belong to $\BB$, since this case in not covered by
 Proposition \ref{prop1BB}(ii). For these $\rho$,   the ideal rectangular filters $H^{LP}_{p,p}$ and $H^{HP}_{p,p}$
 would need more smooth approximations  from  $W_r^d(\R)$   with large enough  $d>1$ to ensure that  $W_r^d(\R) \subset \Lrho$.
 Similarly,   if  $\a\ge 3/2$ then we cannot  defined by $\mu$  as in Theorem \ref{ThFL}(iv).

  \subsection{Linear predictors} 
 \label{SecP}

  We consider the task of recovering of integrals of the future non-observed values of, i.e., 
 convolutions of  underlying process $x$ with 
 anti-causal  kernels, from the observed values $x(s)|_{(-\infty,t]}$, 
 for signals from certain subsets of $\Lrho$.

The following theorem shows that signals with single point spectrum degeneracy feature certain predictability.
This is demonstrated with   predictors 
 explicitly given as causal transfer functions. 
 
 Let  $c>0$ be given. 
For $\w\o\in\R$,  $\nu\in (0,1)$, let \baaa
G(\o,\w\o,\nu)\defi \exp\frac{c}{(\o-\w\o)^2+\nu}.
\eaaa 
Given our choice of $\rho$, we have that, for any $\w\o$ and $\nu$, 
\baaa
 G(\cdot,\w\o,\nu)-1\in \BB,
\eaaa 
i.e, $G(\cdot,\w\o,\nu)\in \CBB$.

Let $\X_{\w\o} $ be the set of all processes  $x\in \Lrho$ such that
\baaa
\|x\|_{\X_{\w\o} }\defi \sup_{\nu\in(0,1)}\| G(\cdot,\w\o,\nu)X\|_{\BB^*}<+\infty, \quad X=\F x.
\eaaa 
We consider $\X_{\w\o} $ as a linear normed space with the corresponding norm.

 Let   $r>0$ be given,  
let 
\baa
\ww H_\g(z)\defi \frac{1}{a-z}\left(
1-\exp\left[-\g\frac{z-a\hphantom{-i}}{z-i\w\o+\g^{-r}}\right]\right),\quad z\in \C,
\label{wK}\eaa
where $\g>0$, and let 
\baa 
\wH_\g(\o)\defi \ww H_\g(i\o),\quad \o\in\R,\qquad \wH_\g=\F^{-1} \wH_\g.
\eaa

 We assume below that $\rho(t)=(1+|t|)^{-\a}$, where $\a\in[0,1/2)$.
Clearly, there exists $r\in(1,2)$ such that $\wH_\g(\cdot)\in W_r^1(\R)\subset\BB$.
Let us select such $r$.
\def\ha{h}\def\Ha{H}
For $a>0$, let $\ha(t)\defi e^{at}\Ind_{t\le 0}$, $t\in\R$, and let $\Ha\defi \F  \ha$. Clearly,
 $\ha\in L_1(\R)$, and  $\Ha(\o)=1/(a-i\o)$ is an anti-causal transfer function.

\begin{theorem}\label{ThP} 
 The functions $\{\wH_\g\}_{\g>0}\subset\BB$ 
 are causal transfer functions defined   on $\Lrho$ such that, for any $\w\o\in \R $, there exists $\oo \g>0$ such that 
\baaa
\sup_{t\in\R}|y(t)-\w y_\g(t)|\le\e\qquad \forall \g\ge \oo \g,
\eaaa
 for  any $x\in\X_{\w\o}$ such that $\|x\|_{\X_{\w\o}}\le 1$. Here
 \baaa
 &&y (t)=\rho(t)\int^{\infty}_t  e^{a(t-s)}x(s),\\
&&\w y_\g (t)=\rho(t)\int_{-\infty}^t h_\g(t-s)x(s)ds,
\eaaa
 where $h_\g\defi\F^{-1}\wH_\g$. 
\end{theorem}
It will be shown in the proof below that the functions  $\wH_\g(\o)$  approximate  $\ww H_a(i\o)=(a-i\o)^{-1}$ for $\o\in \R\setminus D$   as  $\g\to +\infty$.

It can be noted that  functions (\ref{wK}) have been   introduced  in \cite{D21} as predicting transfer functions for signals from $L_2(\R)$ with a single point spectrum degeneracy.

 \section{Proofs}
\label{SecProofs}

  \par
    {\em Proof of Proposition \ref{prop1BB}}.
Let $\w f$, $\w g$, and $\w h$, be the Fourier transforms for $f,g$,and $h$. 
We have that 
\baaa
\|h\|_{\BB}&=&\int_{-\infty}^\infty\rho(t)^{-1} |\w h(t)|dt=
\int_{-\infty}^\infty dt\left|\rho(t)^{-1}\TP\int_{-\infty}^\infty \w f(t-s)\w g(s)ds\right|\\&\le& 
\int_{-\infty}^\infty dt \int_{-\infty}^\infty \rho(t)^{-1} | \w f(t-s)|\,|\w g(s)|ds\le \int_{-\infty}^\infty 
\,|\w g(s)|ds\int_{-\infty}^\infty \rho(t)^{-1} | \w f(t-s)|dt\\ &=
& \int_{-\infty}^\infty 
\,|\w g(s)|ds\int_{-\infty}^\infty  \rho(\tau+s)^{-1}| \w f(\tau)|d\tau
=\int_{-\infty}^\infty \rho(s)^{-1}
\,|\w g(s)|ds\int_{-\infty}^\infty  \rho(\tau)^{-1}| \w f(\tau)|d\tau
\\ &\le& \|f\|_{\BB}\|g\|_{\BB}.
\eaaa We used here the assumptions on $\rho$. This proves statement (i).
\par
Further, let $f\in  W_r^d (\R)$ and $\w f=\F f$.
 We have that $d^mf(\o)/d\o^m\in L_r(\R) $ for $m=0,1,...,k$. 
 By the Hausdorff-Young inequality, we have for $r\in (1,2]$ that and there exists $c_r>0$ such that
\baaa
\left(\int_{-\infty}^\infty (1+|\o|)^{kq}|\w f(\o)|^q d\o)\right)^{1/q} \le c_r\|f\|_{ W_r^d(\R)}.
\eaaa 
We have that   
\baaa
\int_{-\infty}^\infty \rho(\o)|\w f(\o)|d\o
\le \left( \int_{-\infty}^\infty  \rho(\o)^{-r} (1+|\o|)^{-kr} d\o\right)^{1/r}\left(\int_{-\infty}^\infty (1+|\o|)^{kq}|\w f(\o)|^{q}d\o \right)^{1/q}\\
\le C_{d,r,\a}\cdot \|f\|_{ W_r^d (\R)},\eaaa
where
\baaa
C_{k,r,\a}\defi \left(\int_{-\infty}^\infty\frac{ (1+|\o|)^{\a r}}{(1+|\o|)^{dr}}d\o\right)^{1/r}<+\infty,
\eaaa
since $dr-\a r=r(d-\a)>1$ by the choice of $r$. 
This proves statement (ii) and  completes the proof of Proposition \ref{prop1BB}. 
$\Box$
\par
{\em Proof of Proposition \ref{Prop1}}. Let  $U_\BB\defi \{f\in\BB:\ \|f\|_{\BB}\le 1\}$.
For any $f\in U_\BB$, we have that 
\baaa
\left| \langle X_m,f\rangle\right|=\left| \int_{-m}^m x(t)\w f(t)dt\right|\le\esssup_{t:\ |t|\le m}\rho(t) |x(t)| \int_{-m}^m |\rho(t)^{-1}\w f(t)|dt\le \|x\|_{ \Lrho }.
\eaaa
Here $\w f(t)$ is the Fourier transform for $f$.

For $r>0$, let 
$P(r)\defi\{ X\in\BB^*:\  |\langle X,f\rangle| \le r\quad \forall f\in U_\BB\}$.
We have that $\BB$ is a separable Banach space. By the Banach–Alaoglu theorem,  $P(r)$ is sequentially compact in the weak* topology of the dual space $\BB^*$ for any $r>0$; see, e.g., Theorem 3.17 \cite{Rudin}, p.68. Hence 
there exists a sequence  of positive  integers $m_1<m_2<m_3<...\,$ such that
 the subsequence  $\{X_{m_k}\}_{k=1}^\infty$ of the sequence   $\{X_m\}_{m=1}^\infty \subset P(\|x\|_{\Lrho})$ has a  weak* limit in $X\in P(\|x\|_{\Lrho}$. 

Further, for any $f\in U_\BB$ and any integers $n>m>0$, we have that 
\baa\nonumber
\left| \langle X_m-X_n,f\rangle\right|=\left| \int_{t:\ m\le |t|\le n} x(t)\w f(t)dt\right|\le \|x\|_{ \Lrho} \!\!\! \int_{t:\ m\le |t|\le n}\rho(t)^{-1} |\w f(t)|dt \to 0\\ \quad \hbox{as}\quad m\to +\infty.
\label{Cauchy}\eaa

Let us prove that the original sequence $\{X_m\}$ also  has a weak* limit $X$ in $\BB^*$.

Let $f\in\BB$  be given. Let us show that for any  $\e>0$ there exists $N=N(f,\e)$ such that \baa
&&\left| \langle X_{m}-X,f\rangle\right|\le \e\quad \forall k\ge N.
\label{eps}\eaa
Since $X$ is a weak* limit of the 
subsequence  $\{X_{m_k}\}_{k=1}^\infty$, and  by the property 
(\ref{Cauchy}), it follows that for any  $\e>0$ there exists $N=N(f,\e)$ such that \baaa
&&\left| \langle X_{m_k}-X,f\rangle\right|\le \e/2 \quad\forall k\ge N,\\
&&\left| \langle X_{m}-X_{m_k},f\rangle\right|\le \e/2 \quad \forall k\ge N, \ \forall m>m_k.
\eaaa
Hence (\ref{eps}) holds. Hence the sequence $\{X_m\}$  has the same as  $\{X_{m_k}\}_{k=1}^\infty$ weak* limit $X$  in $\BB^*$  that belongs to $P(\|x\|_{\Lrho})$, i.e.,
 $|\langle X,f\rangle|\le \|x\|_{\Lrho}$ for all $f\in\BB$ such that $\|f\|_\BB\le 1$. 
 It follows that  $\|X\|_{\BB^*}\le \|x\|_{\Lrho}$.
\gre{ Furthermore, let a sequence  $\{h_k\}_{k=1}^\infty \subset \Linvrho$ 
 and $s\in\R$  be such that $h_k(t)\ge 0$, $\int_{-\infty}^\infty h_k(t)dt=1$, and
$\lim_{k\to +\infty}|y_{h_k}(s)|=\|x\|_{\Lrho}$. 
We have that $\rho(s)^{-1}|\langle X,g_{h_k}(s,\cdot)\rangle|\to \|x\|_{ \Lrho }$ as $k\to+\infty$. Hence 
$\sup_k|\langle X,g_{h_k}(t,\cdot)\rangle|\ge \|x\|_{\Lrho}$. It follows that the operator norm $\|X\|_{\BB^*}$ is $\|x\|_{\Lrho}$.}
This completes the proof. $\Box$

 {\em Proof  of Lemma \ref{lemma2BB}}. 
By the assumptions on $\rho$, we have that $\rho(\nu)\rho(t)\le \rho(t-\nu)$. Hence
 \baaa
\rho(\nu-t)^{-1}\le\rho(\nu)^{-1}\rho(t)^{-1}. \eaaa
\par  We have that
\baa H(\o) e^{i\o t}= e^{i\o t}\int_{-\infty}^\infty h(\nu) e^{-i \o \nu}d\nu
=\int_{-\infty}^\infty h(\nu)  e^{-i \o (\nu-t)}d\nu=\int_{-\infty}^\infty h(\nu)  e^{i \o (t-\nu)}d\nu\nonumber\\=(h\ast e^{i\o \cdot})(t)
\label{Hehe}
\eaa for $t,\o\in\R$.
In additions, this implies that 
\baaa
H(\o) e^{i\o t}=\int_{-\infty}^\infty h(\nu+t)  e^{-i \o \nu}d\nu.
\eaaa Hence
 \baaa \| H e^{i\cdot t} \|_{\BB}=
\int_{-\infty}^\infty\rho(\nu)^{-1} |h(\nu+t)|d\nu=\int_{-\infty}^\infty\rho(\nu-t)^{-1} |h(\nu)|d\nu
\le \rho(t)^{-1}\int_{-\infty}^\infty\rho(\nu)^{-1} |h(\nu)|d\nu\\=\rho(t)^{-1}\|h\|_{\Linvrho}. 
\eaaa
Further,
\baaa
H(\o) e^{i\o t}-H(\o) e^{i\o s} =\int_{-\infty}^\infty h(\nu+t)-h(\nu+s)  e^{-i \o \nu}d\nu.
 \eaaa
\baaa
\|H e^{i\cdot t}-H e^{i\cdot s} \|_{\BB}=\int_{-\infty}^\infty \rho(\nu)^{-1}|h(\nu+t)-h(\nu+s)|d\nu\to 0\quad \hbox{as}\quad t-s\to 0.
\eaaa
Hence  the function  $H e^{i\cdot t} $ is continuous in $\BB$ with respect to $t\in\R$.
It follows that if the  function $\rho(t)$ is continuous, then the function  $H e^{i\cdot t} g_h(\cdot,t)$ is continuous in $\BB$ with respect to $t\in\R$.
This completes the proof of Lemma \ref{lemma2BB}. 
$\Box$

For  $h\in \Linvrho$,  let  $y_{h,m}(t)\defi\TPR\langle X_m,H e^{i\cdot t}\rangle$,
and   $y_{h}(t)\defi\TPR\langle X,H e^{i\cdot t}\rangle$, where
$x_m(t)\defi x(t)\Ind_{|t|\le m}$, $X_m\defi \F x_m$.

  \par
{{\em Proof  of Proposition \ref{lemmahxy1}}.}  Statement (i) follows from (\ref{Hehe}). Furthermore, the set of convolutions with all
admissible $h$ defines $x$ uniquely, and  Lemma \ref{lemma2BB} ensures the desired estimate.   This completes the proof  of 
Proposition \ref{lemmahxy1}.
  $\Box$

 \par
{\em Proof  of Proposition \ref{lemmahxy}}.  Let  $H\in \BB$ be the Fourier transform of $h\in \Linvrho$.  By Lemma \ref{lemma2BB}, we have that $H e^{i\cdot t}\in \BB$, and
$\rho(t)\|H e^{i\cdot t}\|_{\BB}\le  \|H\|_{\BB}=\|h\|_{\Linvrho}$.

Let as prove statement (i).  We have  that
 \baaa
2\pi\, \sup_t|y_h(t)-y_{h,m}(t)|\le \langle X- X_m, \rho(t) H e^{i\cdot t}\rangle \le \|X- X_m\|_{\BB^*}\sup_t\|\rho(t)H e^{i\cdot t}\|_{\BB}\to 0 
\quad\\ \hbox{as } \quad m\to +\infty.  
 \eaaa
 Hence $y_h(t)=\lim_{m\to +\infty} y_{h,m}(t)= \rho(t)(h\ast x)(t)$ for all $t\in\R$. This proves statement  (i).
 
Let us prove statement (ii). To prove uniqueness of $x$, it suffices to observe that $h\ast x$ is defined for any $h\in \Linvrho$, and 
 if  $(h\ast x)(t)=0$ for all $t\in\R$ then $ x(t)\equiv 0$ a.e..
 
Further, let $B\defi \{h\in \Linvrho:\ \  \|h\|_{\Linvrho}=1\}$. 
We have that $\|x\|_{\Lrho}=\sup_{h\in B}\|y_h\|_{L_\infty(\R)}$. In addition, we have that 
$\|g_h(t,\cdot)\|_{\BB}\le 1$ for any $h\in B$ and $H=\F h$. Hence $|y_h(t)|\le \|X\|_{\BB^*}$  for any $h\in B$. 
Hence  $\|x\|_{\Lrho}\le \|X\|_{\BB^*}$.  This completes the proof  of Proposition \ref{lemmahxy}.
  $\Box$
 \par
 
{\em Proof of Theorem \ref{Th1}}.  We have that  the mappings $\GG:\BB^* \to  \Lrho$ 
and $\F: \Lrho \to \BB^*$ 
are  linear and continuous.  


Let us show that  the mapping $\GG:\BB^*\to  \Lrho $ is injective, i.e. that
\baaa
\hbox{if}\quad x=\GG(X)=0_{ \Lrho }\quad \hbox{then}\quad X=0_{\BB^*}.
\label{injG}\eaaa  
Suppose that  $x=\GG(X)=0_{\Lrho }$. In this case,  \baaa
y_h(t)=\TP\langle X, g_h(\cdot,t) \rangle=0
\eaaa
for all $t\in\R$ and all $h\in\Linvrho$, $H=\F h\in \BB$. 
This means that \baaa
\langle X,f \rangle=0
\eaaa
 for all $f\in\BB$.
   This means that $X=0_{\BB^*}$.  Hence the mapping  $\GG:\BB^*\to  \Lrho$ is injective.

Let us show that $\GG (\F x)=x$ for any $x\in L_\infty$. 
Let $x\in \Lrho $, $h\in \Linvrho$, and $m>0$. Let $X=\F x$, $x_m(t)=x(t)\Ind _{|t|\le m}$, 
$X_m=\F x_m=\int_{-m}^m e^{-i\o t}x(t)$, and $y_m=h\ast x_m$.
We know from that properties of the Fourier transform that
$x_m=\GG X_m$.       

Suppose that $\w x\defi \GG(X)\neq x$. 
Let $\w x_m(t)=\w x(t)\Ind _{|t|\le m}$ and $\w  y_m(t)=(h\ast \w x_m)(t)$. 
By the definitions,  $\w y_m(t) 
=\TP\langle X_m,g_h(t,\cdot )\rangle =y_m(t)$ for any $t\in\R$.  Hence 
\baaa \rho(t)(h\ast x)(t)=\lim_{m\to +\infty}\TPR\langle X_m,\g_h(t,\o))\rangle=\TPR\langle X,H e^{i\cdot t} \rangle 
\eaaa
for any $t\in\R$ and $m>0$. Since it holds for all $h$, it follows that $\w x_m(t)=x_m(t)$ for all $t$ and all $m>0$.
Hence $x=\GG X$.   In its turn, this implies that   $\F \GG (\F x)=\F x$, and   $\F (\GG X)=X$ for $X=\F x$, and
$\F (\GG X)=X$ for all  $X\in\BB^*$.  

Hence   the mapping $\GG: \BB^*\to\Lrho$  is a bijection,  $\GG^{-1}=\F$, and $\F^{-1}=\GG$. 
  
The continuity of the mapping $\GG: \BB^*\to\Lrho$ is obvious. The continuity of the mapping  $\F=\GG^{-1}: \Lrho \to \BB^*$ 
follows from Proposition \ref{Prop1}; alternatively,  it can be shown  using e.g. Corollary 2.12(c) in \cite{Rudin}, p. 49.
This completes the proof  of Theorem \ref{Th1}. $\Box$

{\em Proof of Corollary  \ref{propP}}.  Let $h(t)=y(-t)$ and $H= \F \oo h$, i.e., 
\baaa
H(\o)=\int_{-\infty}^\infty e^{-i\o t}\oo h(t)dt=\int_{-\infty}^\infty e^{-i\o t}\oo y(-t)dt=\int_{-\infty}^\infty e^{i\o s}\oo y(s)ds
\int_{-\infty}^\infty \overline{e^{-i\o s}}\oo y(s)ds=\oo Y(\o).
\eaaa 
 Hence 
 \baaa  \int_{-\infty}^\infty x(t)\oo y(t)dt=\int_{-\infty}^\infty x(t)\oo h(-t)dt=(\oo h\ast x)(0)=
 \TP\langle X, H e^{i\cdot 0} \rangle
 =\TP\langle X, H \rangle =\TP\langle X,\oo Y \rangle.
\eaaa
 $\Box$

{\em Proof of Proposition \ref{propTF}}. 
Let $x_m(t)=\Ind_{\{|t|\le m\}}$, $m=1,2,...$,  and $X_m=\F x_m$, $y_m=h\ast x_m$.
By Proposition \ref{lemmahxy1},
we have that 
 \baaa
y_m(t)=(h\ast x_m) (t) =\TP\langle X_m, H e^{i\cdot t} \rangle, \quad y(t)=(h\ast x) (t) =\TP\langle X, H e^{i\cdot t} \rangle\quad \hbox{for}\quad t\in\R.
 \eaaa
 It was shown above that $X_m\to X$ weakly in $\BB^*$, and $H e^{i\cdot t}\in \BB$. 
  Clearly,  $y_m(t)\to y(t)$ as $m\to +\infty$. Then the proof follows.
 $\Box$  

 \par
  {\em Proof of Theorem \ref{ThCBL}}.  Let   $p\in (\O,+\infty)$ and  $H\in\Linvrho$ be selected such that
 $H(\o)=1$ for $\o\in[-\O,\O]$, that  $H(\o)=0$ if $|\o|>p$, and that $H$ has bounded and continuous derivatives $d^kH(\o)/d\o^k$ for any integer $k>0$. 
 Let $X=\F x$ and  $h=\F^{-1}H$.   Clearly, $h\in\Linvrho$  and $X=HX$. Hence $x=h\ast x$. Therefore, $x$ also has continuous derivatives of any order.
 $\Box$

\comm{\par Only W^1:
  {\em Proof of Theorem \ref{ThCBL}}.  Let  $p,q\in\R$ be such that $q>p>\O$. Let $x\in\Lrho$, and let  $H^{LP}_{p,q}$ be defined by (\ref{LPF}).
 Let $X=\F x$ and  $h^{LP}=\F^{-1}H^{LP}_{p,q}$.   Clearly, $X=H^{LP}_{p,q}X$ and  $h^{LP}\in\Linvrho$. Hence $x=h^{LP}\ast x$. Therefore, this signal is continuous.
 $\Box$}

  \par
{\em Proof of Theorem \ref{ThC}}.   The assumptions on $H$ imply that
$H(i\o)=a+\int_{0}^\infty h(t) e^{-i\o t}dt$ for $\o\in\R$ and $h\in\Linvrho$. Then the result follows from  Lemma \ref{lemmahxy}(i). $\Box$
 
In addition,  let us provide an alternative  proof of  Theorem \ref{ThC} that does not rely on Lemma \ref{lemmahxy}. {  
It suffices to consider $X\in \BB^*$.Let  $x=\F^{-1} X$ be such that $x(t)=0$ for $t<\tau$. For $m>0$, let 
 $x_m(t)=x(t)\Ind_{|t|\le m}$ and $X_m=\F^{-1} x_m=\int_{-m}^m e^{-i\cdot t} x(t)dt$.
   Further, let $\w X_m\defi H(\cdot/i) X_m$ and $\w x_m\defi \F^{-1}\w X_m$.
 From the standard theory of causal transfer functions for processes from $L_2(\R)$, we know that
 $\w x_m(\tau)=0$ for all $m$. As was shown in the proof of Theorem \ref{Th1}, we have that $X_m\to X$ as $m\to +\infty$ in weak* topology of $\BB^*$. 
 Since $H(\cdot/i)f\in\BB$ for any $f\in\BB$, it follows that  $\w X_m =H(\cdot/i) X_m \to \w X$ as $m\to +\infty$ in weak* topology of $\BB^*$.  
 Hence $\w x_m(\tau)\to \w x(\tau)$ as $m\to +\infty$. Therefore, $\w x(\tau)=0$. This completes the proof.
 $\Box$
 
{\em Proof of Theorem \ref{ThFL}}. Since $ W_r^d(\R)\subset \BB$ with selected $d,r$, the functions  $H^{LP}_{p,q}$ belong to $\BB$,
 by Proposition \ref{prop1BB} (ii). Hence s  $H^{HP}_{p,q}\in \CBB$.
 
Let  $D= (-p,p)$  and  let  $H=H^{LP}_{p,q}$, $h=\F^{-1}H$. Let $x\in  \Lrho$,
$y(t)=(h\ast x)(t)$, and $Y=\F y$.  We have that
 \baaa
 \int_{-\infty}^\infty y(t)f(t)dt=
  \TP\langle  H X,F\rangle =\TP\langle  X,HF\rangle=0\eaaa 
  for any $f\in \Linvrho$ such  that $F|_{ \R \setminus D}\equiv 0$, since $HF=0$ in this case.
  By Definition \ref{defGap}, $y$ has a spectrum gap $D$.
  This completes the proof for the case (i) where $H=H^{LP}_{p,q}$. The case (ii)  for $H=H^{HP}_{p,q}$
  is similar. To prove statements (iii)-(iv), one can verify directly that $\mu$ satisfies the required conditions. 
  $\Box$
  \par
{\em Proof of Theorem \ref{ThP}}. 
 Let \baaa
U_\g(z)\defi \exp\left[-\g\frac{a-z\,\,\,}{z-i\w\o+ \g^{-r}}\right],
 \qquad z\in\C.
 \eaaa
 We have that  \baaa
\ww H_\g(z)=   \frac{1}{a-z}\left(1-U_\g(z)\right),
 \qquad z\in\C.
 \eaaa
Clearly, $1-\exp(z)=-\sum_{k=1}^{+\infty}(-1)^k z^k/k!$ for $z\in\C$.  
Hence  \baaa
\ww H_\g(z) =-\sum_{k=1}^{+\infty}\frac{(-1)^k\g^k(z-a)^k}{k!(z-i\w\o+\g^{-r})^k} \brea
=-(z-a)\sum_{k=1}^{+\infty}\frac{(-1)^k\g^k(z-a)^{k-1}}{k!(z-i\w\o+\g^{-r})^k}.
\eaaa
Therefore, the pole at $z=a$ of $(a-z)^{-1}$  is being compensated by multiplying on $1-U_\g(z)$.
Hence functions $\ww H_\g(z)$ are
 continuous on $\oo\C^+$ and analytic on $\C^+$. 
Clearly,  $\wH_\g(\o)=\ww H_\g(\cdot/i)\in  W_r^1 (\R)\subset\BB$.
This means that
$\wH_\g$ are causal transfer functions belonging to  $\BB$.   Then
statement (i) follows.

\def\Ko{U}

 Let \baaa
V_\g(\o)\defi \wH_\g(i\o)-H(i\o)= \frac{U_\g(i\o)}{a-i\o},
 \qquad \o\in\R.
 \eaaa

\def\ew{(\cdot i)}
\begin{lemma}\label{lemmaV}
$\|V_\g(i\cdot)G(\cdot,\w\o,\g^{-r})^{-1}\|_{\BB}\to 0$ as $\g\to+\infty$.
\end{lemma}
\par
{\em Proof of Lemma \ref{lemmaV}.}
We have that $V_\g\ew G(\o,\w\o,\g^{-r})^{-1}=e^{-\psi_\g(\o)}$, where
 \baaa
\psi_\g(\o)\defi \frac{c}{(\o-\w\o)^2+\g^{-r}}+\g\frac{a-i\o}{i(\o-\w\o)+\g^{-r}}.
\eaaa
The proof for Lemma \ref{lemmaV}  explores the fact that $\Re\psi_\g(\o)\to +\infty$ as $\g\to +\infty$ for all 
$\o\neq \w\o$ and  $\inf_\g\Re\psi_\g(\o)$  is bounded  
in a neighbourhood of $\o=\w\o$; it is rather technical and will be omitted. This gives that  $\|V_\g(i\cdot)G(\cdot,\w\o,\g^{-r})^{-1}\|_{W_r^1(\R)}\to 0$  and hence 
 $\|V_\g(i\cdot)G(\cdot,\w\o,\g^{-r})^{-1}\|_{\BB}\to 0$ as $\g\to+\infty$. The proof 
 is  similar to the proof of Lemma 7.1 in \cite{D21} and is rather technical; it will be omitted here.
$\Box$

Further, we have that 
\baaa
2\pi (y(t)-\w y_d(t))&=& \rho(t)\langle  X,[H_a-\wH_\g]e^{i\cdot t}\rangle
= \rho(t)\langle   X, \, e^{i \cdot t}V_\g(\cdot,t)\rangle\\&
=&\rho(t)\langle   X,\,G(\cdot,\w\o,\g^{-r})G(\cdot,\w\o,\g^{-r})^{-1}e^{i \cdot t}V_\g(\cdot,t)\rangle
\\&=&\rho(t)\langle   G(\cdot,\w\o,\g^{-r})X,\,G(\cdot,\w\o,\g^{-r})^{-1}e^{i \cdot t}V_\g(\cdot,t)\rangle.
\eaaa
By Lemma \ref{lemma2BB}, we have that \baaa
\left\|\rho(t)e^{i \cdot t}G(\cdot,\w\o,\g^{-r})^{-1}V_\g\right\|_{\BB}\le \left\|e^{i \cdot t}G(\cdot,\w\o,\g^{-r})^{-1}V_\g\right\|_{\BB}.
\eaaa
 Hence 
\baaa
2\pi|y(t)-\w y(t)|&\le& \rho(t)\left\|G(\cdot,\w\o,\g^{-r})X\right\|_{\BB^*} 
\left\|e^{i \cdot t}G(\cdot,\w\o,\g^{-r})^{-1}V_\g\right\|_{\BB}\\ 
&\le &\left\|G(\cdot,\w\o,\g^{-r})X\right\|_{\BB^*} 
\left\|G(\cdot,\w\o,\g^{-r})^{-1}V_\g\right\|_{\BB}\\
&\le&\sup_{\g>0}\|G(\cdot,\w\o,\g^{-r})X\|_{\BB^*}\| G(\cdot,\w\o,\g^{-r})^{-1}V_\g\|_{\BB}\to 0\quad \hbox{as}\quad \g\to +\infty.
\eaaa
By Lemma \ref{lemmaV}, the proof of Theorem  \ref{ThP} follows. $\Box$

\section{Concluding remarks and further research}
 \begin{enumerate}
 \item Theorem \ref{ThP} implies that a  signals from
 $\Lrho$ covered by this theorem are uniquely defined by their past history.  
 Another  implication of Theorem \ref{ThP} 
 is  that the impact of the signal's history on  its future values  diminishes with time  for  unbounded signals from
 $\Lrho$ with single point spectrum degeneracy. This is because  the predicting kernels in  Theorem \ref{ThP} belong to $L_r(\R)$ for some $r>1$, 
 belong to $L_r(\R)$ for some $r>1$.
 
   \item
   So far, it is unclear if the set of all  bandlimited signals 
 is everywhere dense in  the space  $\Lrho$, similarly to the space $L_2(\R)$, where the set of all band-limited processes is everywhere dense. 
 \item
  It is unclear how describe spectrum gaps with more general choice of $\rho$, such as, for example, $\rho(t)=e^{-\a|t|}$, where  $\a>0$.
\end{enumerate}

\end{document}